\documentclass[a4paper,twoside]{article}

\usepackage{epsfig}
\usepackage{subcaption}
\usepackage{calc}
\usepackage{amssymb}
\usepackage{amstext}
\usepackage{amsmath}
\usepackage{amsthm}
\usepackage{multicol}
\usepackage{pslatex}
\usepackage{apalike}
\usepackage{algorithm2e}
\usepackage[bottom]{footmisc}

\usepackage{url}
\usepackage{xcolor}
\usepackage{adjustbox}

\usepackage{SCITEPRESS}     

\begin{document}

\title{QuLTSF: Long-Term Time Series Forecasting \\ with Quantum Machine Learning}

\author{\authorname{Hari Hara Suthan Chittoor\sup{1}\orcidAuthor{0000-0003-1363-606X}, Paul Robert Griffin\sup{1}\orcidAuthor{0000-0003-2294-5980}, Ariel Neufeld\sup{2}\orcidAuthor{0000-0001-5500-5245}, Jayne Thompson\sup{3,4}\orcidAuthor{0000-0002-3746-244X}, Mile Gu\sup{2,5}\orcidAuthor{0000-0002-5459-4313}}
\affiliation{\sup{1}School of Computing and Information Systems, Singapore Management University, Singapore, 178902}
\affiliation{\sup{2}Nanyang Quantum Hub, School of Physical and Mathematical Sciences, Nanyang Technological University, Singapore}
\affiliation{\sup{3}Institute of High Performance Computing (IHPC), Agency for Science, Technology and Research (A*STAR), Singapore}
\affiliation{\sup{4}Horizon Quantum, 29 Media Cir., \#05-22/23/24 ALICE@MEDIAPOLIS, South Tower, Singapore
138565}
\affiliation{\sup{5}Centre for Quantum Technologies, National University of Singapore, Singapore}
\email{\{haric, paulgriffin\}@smu.edu.sg, ariel.neufeld@ntu.edu.sg,
thompson.jayne2@gmail.com, 
mgu@quantumcomplexity.org} 
}


\abstract{Long-term time series forecasting (LTSF) involves predicting a large number of future values of a time series based on the past values. This is an essential task in a wide range of domains including weather forecasting, stock market analysis and disease outbreak prediction. Over the decades LTSF algorithms have transitioned from statistical models to deep learning models like transformer models. Despite the complex architecture of transformer based LTSF models `Are Transformers Effective for Time Series Forecasting? (Zeng et al., 2023)' showed that simple linear models can outperform the state-of-the-art transformer based LTSF models. Recently, quantum machine learning (QML) is evolving as a domain to enhance the capabilities of classical machine learning models. In this paper we initiate the application of QML to LTSF problems by proposing \textit{QuLTSF}, a simple hybrid QML model for multivariate LTSF. Through extensive experiments on a widely used weather dataset we show the advantages of QuLTSF over the state-of-the-art classical linear models, in terms of reduced mean squared error and mean absolute error.
}

\onecolumn \maketitle \normalsize \setcounter{footnote}{0} \vfill

\section{\uppercase{Introduction}}
\label{sec:introduction}

Time series forecasting (TSF) is the process of predicting future values of a variable using its historical data. TSF is an import problem in many fields like weather forecasting, finance, power management etc. There are broadly two approaches to handle TSF problems: statistical models and deep learning models. Statistical models, like ARIMA, are the traditional work horse for TSF since the $1970$'s \cite{Book_hyndman2018forecasting,Book_hamilton2020time_serires_analsis}. Deep learning models, like recurrent neural networks (RNN's), often outperform statistical models in large-scale datasets \cite{Timeseries_forecasting_with_deep_learning_a_survey_Lim_2021}. 

Increasing the prediction horizon strain's the models predictive capacity. The prediction length of more than $48$ future points is generally considered as \textit{long-term time series forecasting} (LTSF) \cite{zhou2021informer}. Transformers and attention mechanism proposed in \cite{vaswani2017attention_is_all_you_need} gained a lot of attraction to model sequence data like language, speech etc. There is a surge in the application of transformers to LTSF leading to several time series transformer models \cite{zhou2021informer,wu2021autoformer,zhou2022fedformer,liu2022pyraformer,wen2023transformers_in_time_series_survey}. Despite the complicated design of transformer based models for LTSF problems, \cite{zeng2023_are_transformers_effective_for_time_series_forecasting} showed that a simple autoregressive model with a linear fully connected layer can outperform the state-of-the-art  transformer models.

\textit{Quantum machine learning} (QML) is an emerging field that combines quantum computing and machine learning to enhance tasks like classification, regression, generative modeling etc., using the currently available \textit{noisy intermediate-scale quantum} (NISQ) computers \cite{preskill2018quantum,book_schuld2021machine,Book_QML_Osvaldo}. Hybrid models containing classical neural networks and variational quantum circuits (VQC's) are increasingly becoming popular for various machine learning tasks, thanks to rapidly evolving software tools \cite{bergholm2018pennylane,broughton2020tensorflow_quantum}. The existing QML models for TSF focus on RNN's \cite{emmanoulopoulos2022quantum,Hybrid_Quantum_Classical_RNN_for_TSF}. In time series analysis, recurrent quantum circuits have demonstrated provable computational and memory advantage during inference, however learning such models remains challenging at scale \cite{binder2018practical_Mile_Jayne}. 


In this paper we initiate the application of QML to LTSF by proposing \textit{QuLTSF} a simple hybrid QML model. QuLTSF is a combination of classical linear neural networks and VQC's. Through extensive experiments, on the widely used weather dataset, we show that the QuLTSF model outperforms the state-of-the-art linear models proposed in \cite{zeng2023_are_transformers_effective_for_time_series_forecasting}.

\textbf{Organization of the paper:} Section $2$ provides background information on quantum computing and quantum machine learning. Section $3$ discusses the problem formulation and related work. In Section $4$ we introduce our QuLTSF model. Experimental details, results, discussion and future research directions are given in Section $5$ and Section $6$ concludes the paper.

\section{\uppercase{Background}}

\subsection{Quantum Computing}

The fundamental unit of quantum information and computing is the qubit. In contrast to a classical bit, which exist in either $0$ or $1$ state, the state of a qubit can be $0$ or $1$ or a superposition of both. In the Dirac's ket notation the state of a qubit is given as a two-dimensional amplitude vector
\begin{equation*}
    |\psi\rangle = \begin{bmatrix} \alpha_0  \\ \alpha_1  \end{bmatrix} = \alpha_0 |0\rangle + \alpha_1 |1\rangle,
\end{equation*}
where $\alpha_0$ and $\alpha_1$ are complex numbers satisfying the unitary norm condition, i.e., $|\alpha_0|^2 + |\alpha_1|^2 = 1 $. The state of a qubit is only accessible through measurements. A measurement in the computational basis collapses the state $|\psi \rangle$ to a classical bit $x \in \{0,1\}$ with probability $|\alpha_x|^2$. A qubit can be transformed from one state to another via reversible unitary operations also known as quantum gates \cite{nielsen2010quantum}.

Shor's algorithm \cite{shors_algorithm_1994} and Grover's algorithm \cite{grover1996fast} revolutionized quantum algorithms research by providing theoretical quantum speedups compared to classical algorithms. The implementation of these algorithms require larger number of qubits with good error correction \cite{book_lidar2013quantum_error_correction}. However, the current available quantum devices are far from this and often referred to as the noisy intermediate-scale quantum (NISQ) devices \cite{preskill2018quantum}. Quantum machine learning (QML) is an emerging field to make best use of NISQ devices \cite{book_schuld2021machine,Book_QML_Osvaldo}.

\subsection{Quantum Machine Learning}
The most common QML paradigm refers to a two step methodology consisting of a \textit{variational quantum circuit} (VQC) or \textit{ansatz} and a classical optimizer, where VQC is composed of parametrized quantum gates and fixed entangling gates. The classical data is first encoded into a quantum state, using a suitable data embedding procedure like amplitude embedding, angle embedding etc. Then, the VQC applies a parametrized unitary operation which can be controlled by altering its parameters. The output of the quantum circuit is given by measuring the qubits. The parameters of the VQC are optimized using classical optimization tools to minimize a predefined loss function. Often VQC's are paired  with classical neural networks, creating hybrid QML models. Several packages, for instance \cite{bergholm2018pennylane,broughton2020tensorflow_quantum}, provide software tools to efficiently compute gradients for these hybrid QML models. 

\section{\uppercase{Preliminaries}}

\subsection{Problem Formulation}
Consider a multivariate time series dataset with $M$ variates. Let $L$ be the size of the look back window or sequence length and $T$ be the size of the forecast window or prediction length. Given data at $L$ time stamps $\mathbf{x}_{1:L} = \{\mathbf{x}_1, \cdots, \mathbf{x}_L \} \in \mathbb{R}^{L \times M}$, we would like to predict the data at future $T$ time stamps $\mathbf{\hat{x}}_{L+1:L+T} = \{\mathbf{\hat{x}}_{L+1}, \cdots, \mathbf{\hat{x}}_{L+T} \} \in \mathbb{R}^{T \times M}$ using a QML model. Predicting more than $48$ future time steps is typically considered as Long-term time series forecasting (LTSF) \cite{zhou2021informer}. We consider the channel-independence condition where each of the univariate time series data at $L$ time stamps $\mathbf{x}^{m}_{1:L} = \{x^{m}_1, \cdots, x^{m}_L \} \in \mathbb{R}^{L \times 1}$ is fed separately into the model to predict the data at future $T$ time stamps $\mathbf{\hat{x}}^{m}_{L+1:L+T} = \{\hat{x}^{m}_{L+1}, \cdots, \hat{x}^{m}_{L+T} \} \in \mathbb{R}^{T \times 1}$, where $m \in \{1,\cdots, M\}$. To measure discrepancy between the ground truth and prediction, we use Mean Squared Error (MSE) loss function defined as

\begin{equation}
\label{eq MSE}
    \text{MSE} = \mathbb{E}_{\mathbf{x}} \Biggr[\frac{1}{M} \sum_{m=1}^{M} \|  \mathbf{x}^{m}_{L+1:L+T} -\mathbf{\hat{x}}^{m}_{L+1:L+T} \|_2^2 \Biggr].
\end{equation}

\subsection{Related Work}

LTSF is an extensive area of research. In this section, we provide a concise overview of works most relevant to our problem formulation. Since the introduction of transformers \cite{vaswani2017attention_is_all_you_need}, there has been an increase in
transformer based models for LTSF \cite{wu2021autoformer,zhou2021informer,liu2022pyraformer,zhou2022fedformer,LogsparseTrans_li2019enhancing}. \cite{wen2023transformers_in_time_series_survey} provided a comprehensive survey on transformer based LTSF models. Despite the sophisticated architecture of transformer based LTSF models, \cite{zeng2023_are_transformers_effective_for_time_series_forecasting} showed that simple linear models can achieve superior performance compared to the state-of-the-art transformer based LTSF models.

\cite{zeng2023_are_transformers_effective_for_time_series_forecasting} proposed three models: Linear, NLinear and DLinear. Linear model is just a one layer linear neural network. In NLinear, the last value of the input is subtracted before being passed through the linear layer and the subtracted part is added back to the output. DLinear first decomposes the time series into trend, by a moving average kernel, and seasonal components which is a famous method in time series forecasting \cite{Book_hamilton2020time_serires_analsis} and is extensively used in the literature \cite{wu2021autoformer,zhou2022fedformer,zeng2023_are_transformers_effective_for_time_series_forecasting}. Two similar but distinct linear models are trained for trend and seasonal components. Adding the outputs of these two models gives the final prediction. 
We adapt the simple Linear model to propose our QML model in the next section.

\section{{QuLTSF}}

\begin{figure}[!h]
  \vspace{-0.2cm}
  \centering
   {\epsfig{file = 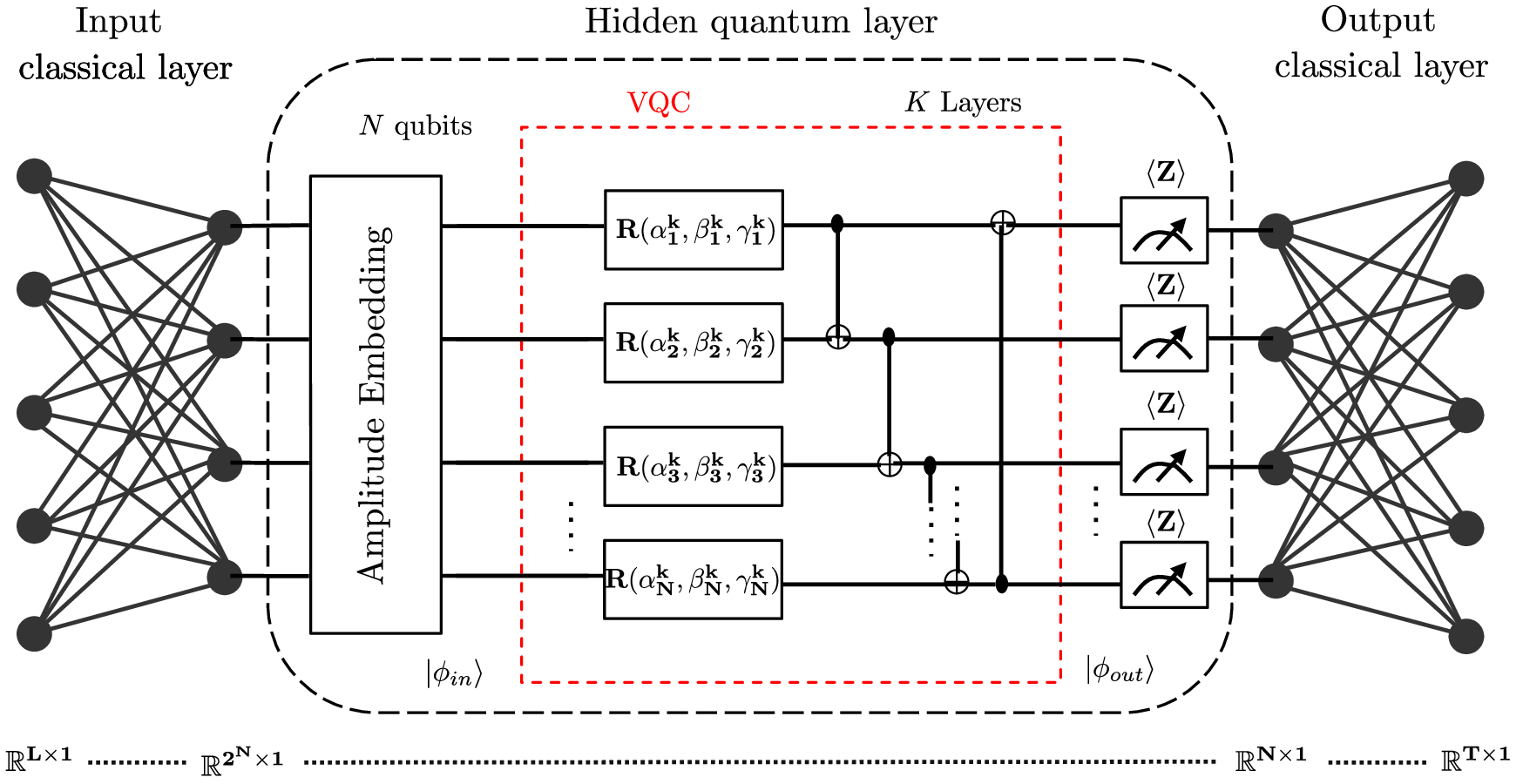, width = 7.5cm}}
  \caption{QuLTSF model architecture.}
  \label{fig:QDLinear architecture}
\end{figure}

In this section, we propose \textit{QuLTSF} a hybrid QML model for LTSF and is illustrated in Fig. \ref{fig:QDLinear architecture}. It is a hybrid model consisting of input classical layer, hidden quantum layer and an output classical layer. The input classical layer maps the $L$ input features in to a $2^N$ length vector. Specifically, the input sequence $\mathbf{x}^{m}_{1:L} \in \mathbb{R}^{L \times 1}$ is given to the input classical layer with trainable weights $\mathbf{W}_{in} \in \mathbf{R}^{2^N \times L}$ and bias $\mathbf{b}_{in} \in \mathbf{R}^{2^N \times 1}$, and it outputs a $2^N$ length vector
\begin{equation}
    \mathbf{y}_{1} = \mathbf{W}_{in} \mathbf{x}^{m}_{1:L}  + b_{in}.
\end{equation}
The output of the input classical layer, $y_1 \in \mathbf{R}^{2^N \times 1}$, is given as input to the hidden quantum layer which consists of $N$ qubits. We use amplitude embedding \cite{book_schuld2021machine} to encode $2^N$ real numbers in $y_1$ to a quantum state $|\phi_{in} \rangle$. We use hardware efficient ansatz \cite{Book_QML_Osvaldo} as a VQC, and is composed of $K$ layers each containing a trainable parametrized single qubit gate on each qubit and a fixed circular entangling circuit with CNOT gates as shown in Fig. \ref{fig:QDLinear architecture}. Every single qubit gate has $3$ parameters and the total number of parameters in $K$ layers is $3NK$. The output of VQC is given as
\begin{equation}
    | \phi_{out} \rangle = (\text{VQC}) | \phi_{in}\rangle.
\end{equation}
We consider the expectation value of Pauli-$Z$ observable for each qubit, which serves as the output of hidden quantum layer and is denoted as $y_2 \in \mathbb{R}^{N \times 1}$.
Finally, $y_2$ is passed through output classical layer with trainable weights $\mathbf{W}_{out} \in \mathbf{R}^{T \times N}$ and bias $\mathbf{b}_{out} \in \mathbf{R}^{T \times 1}$, which maps $N$ length quantum hidden layer output to predicted $T$ length vector
\begin{equation}
   \mathbf{\hat{x}}^{m}_{L+1:L+T} = \mathbf{W}_{out} \mathbf{y}_{2} + b_{out}.
\end{equation}
The parameters of the hidden quantum layer and two classical layers can be jointly trained, similar to classical machine learning, using software packages like PennyLane \cite{bergholm2018pennylane}.

\section{\uppercase{Experiments}}

In this section, we validate the superiority of the proposed QuLTSF model through extensive experiments. The code for experiments is publicly available on GitHub\footnote{https://github.com/chariharasuthan/QuLTSF}. 
All experiments are conducted on SMU's Crimson GPU cluster\footnote{https://violet.scis.dev/}.


\begin{table*}[htbp]

\caption{Multivariate long-term time series forecasting (LTSF) results in terms of MSE and MAE between the proposed QuLTSF model and the state-of-the-art on the widely used weather dataset. Sequence length $L=336$ and prediction length $T \in \{96, 192, 336, 720\}$. The best results are in \textbf{bold} and the second best results are \underline{underlined}.}
\label{table for MSE MAE comparison with SOTA}

\setlength{\tabcolsep}{6pt} 
\renewcommand{\arraystretch}{1.4}
\centering

\begin{adjustbox}{width=\textwidth}
\begin{tabular}{|c||c|c||c|c||c|c||c|c||c|c||c|c||c|c|}
\hline

\textbf{Methods} & \multicolumn{2}{|c||}{\textbf{QuLTSF*}} & \multicolumn{2}{|c||}{\textbf{Linear}} & \multicolumn{2}{|c||}{\textbf{NLinear}} & \multicolumn{2}{|c||}{\textbf{DLinear}} & \multicolumn{2}{|c||}{\textbf{FEDformer}} & \multicolumn{2}{|c||}{\textbf{Autoformer}} & \multicolumn{2}{|c|}{\textbf{Informer}}  \\

\hline

\textbf{Metric} & MSE & MAE & MSE & MAE & MSE & MAE & MSE & MAE & MSE & MAE & MSE & MAE & MSE & MAE \\
\hline \hline
$96$ & $\mathbf{0.156}$  & $\mathbf{0.211}$ & $\underline{0.176}$  & $0.236$  & $0.182$  & $\underline{0.232}$  & $\underline{0.176}$  & $0.237$  & $0.217$  & $0.296$  & $0.266$  & $0.336$ & $0.300$ & $0.384$ \\
\hline

$192$ & $\mathbf{0.199}$  & $\mathbf{0.253}$ & $\underline{0.218}$  & $0.276$  & $0.225$  & $\underline{0.269}$  & $0.220$  & $0.282$  & $0.276$  & $0.336$  & $0.307$  & $0.367$ & $0.598$ & $0.544$ \\
\hline

$336$ & $\mathbf{0.248}$ & $\mathbf{0.296}$ & $\underline{0.262}$ & $0.312$  & $0.271$  & $\underline{0.301}$  & $0.265$  & $0.319$  & $0.339$  & $0.380$  & $0.359$  & $0.395$ & $0.578$ & $0.523$ \\
\hline

$720$ & $\textbf{0.315}$  & $\textbf{0.346}$ & $0.326$  & $0.365$  & $0.338$  & $\underline{0.348}$  & $\underline{0.323}$ & $0.362$  & $0.403$  & $0.428$  & $0.419$  & $0.428$ & $1.059$ & $0.741$ \\
\hline

\end{tabular}
\end{adjustbox}

*QuLTSF is implemented by us; Other results are from \cite{zeng2023_are_transformers_effective_for_time_series_forecasting}.
\end{table*}


\subsection{Dataset Description}
We evaluate the performance of our proposed QuLTSF model on the widely used Weather dataset. It is recorded by Max-Planck Institute of Biogeochemistry\footnote{\url{https://www.bgc-jena.mpg.de/wetter/}} and consists of $21$ meteorological and environmental features like air temperature, humidity, carbon dioxide concentration in parts per million etc. This is recorded in 2020 with granularity of 10 minutes and contains $52,696$ timestamps. $70$ percent of the available data is used for training, $20$ percent for testing and the remaining data for validation.

\subsection{Baselines}

We choose all three state-of-the-art linear models namely Linear, NLinear and DLinear proposed in \cite{zeng2023_are_transformers_effective_for_time_series_forecasting} as the main baselines. We also consider a few transformer based LTSF models FEDformer \cite{zhou2022fedformer}, Autoformer \cite{wu2021autoformer}, Informer \cite{zhou2021informer} as other baselines.
Moreover \cite{wu2021autoformer,zhou2021informer} showed that transformer based models outperform traditional statistical models like ARIMA \cite{book_box2015time_series_analysis} and other deep learning based models like LSTM \cite{bai2018empirical} and DeepAR \cite{salinas2020deepar}, thus we do not include them in our baselines.

\subsection{Evaluation Metrics}
Following common practice, in the state-of-the-art we use MSE (\ref{eq MSE}) and Mean Absolute
Error (MAE) (\ref{eq MAE}) as evaluation metrics
\begin{equation}
\label{eq MAE}
    \text{MAE} = \mathbb{E}_{\mathbf{x}} \Biggr[\frac{1}{M} \sum_{m=1}^{M} ||  \mathbf{x}^{m}_{L+1:L+T} -\mathbf{\hat{x}}^{m}_{L+1:L+T} ||_{1} \Biggr].
\end{equation}

\subsection{Hyperparameters}

The number of qubits $N=10$ and number of VQC layers $K=3$ in the hidden quantum layer. Adam optimizer \cite{kingma2017adammethodstochasticoptimization} is used to train the model in order to minimize the MSE over the training set. Batch size is $16$ and initial learning rate is $0.0001$. For more hyperparameters refer to our code. 


\subsection{Results}

\begin{figure}[htbp]
  \centering
   {\epsfig{file = 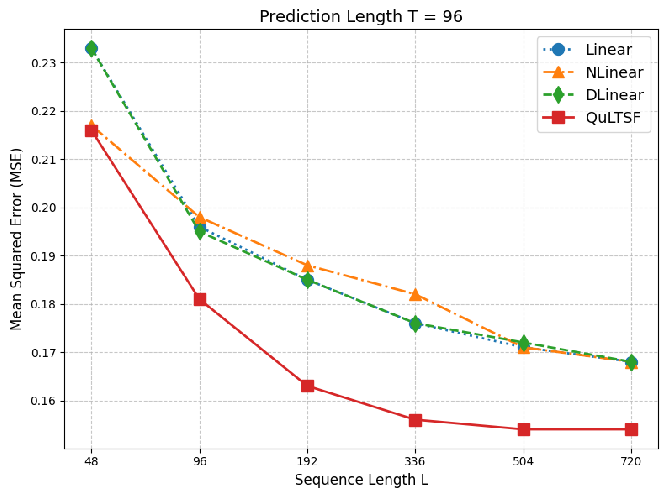, width = 7.6cm}}
  \caption{MSE comparison with fixed prediction length $T=96$ and varying sequence length $L\in \{48, 96, 192, 336, 504, 720\}$.}
  \label{fig:pred length 96}
\end{figure}

\begin{figure}[htbp]
  \centering
   {\epsfig{file = 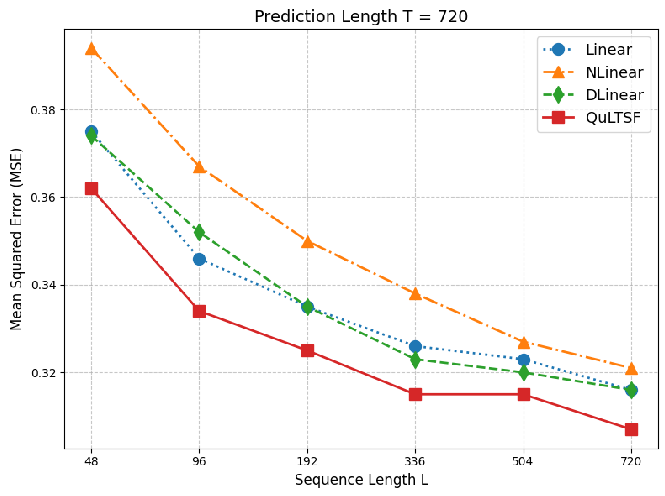, width = 7.6cm}}
  \caption{MSE comparison with fixed prediction length $T=720$ and varying sequence length $L\in \{48, 96, 192, 336, 504, 720\}$.}
  \label{fig:pred length 720}
\end{figure}

For a fair comparison, we choose the fixed sequence length $L=336$ and $4$ different prediction lengths $T\in \{96, 192, 336, 720\}$ as in \cite{zeng2023_are_transformers_effective_for_time_series_forecasting,zhou2022fedformer,wu2021autoformer,zhou2021informer}.
Table \ref{table for MSE MAE comparison with SOTA} provides comparison of MSE and MAE of QuLTSF with $6$ baselines. 
The best and second best results are highlighted in \textbf{bold} and \underline{underlined} respectively. Our proposed QuLTSF outperform all the baseline models in all $4$ cases.

To further validate QuLTSF against the baseline linear models we conduct experiments for the QuLTSF and classical models with varying sequence lengths $L \in \{48, 96, 192, 336, 504, 720\}$ and plot the MSE results for a fixed smaller prediction length $T=96$ in Fig. \ref{fig:pred length 96}, and for a fixed larger prediction length $T=720$ in Fig. \ref{fig:pred length 720}. In all cases QuLTSF outperforms all the baseline linear models.

\subsection{Discussion and Future Work}

QuLTSF uses generic hardware-efficient ansatz. Similar to classical machine learning in QML we need to choose the ansatz, if possible, based on dataset and domain expertise. Searching for optimal ansatz is a research direction by itself \cite{du2022quantum}. Finding better ans\"atze for QML based LTSF models for different datasets is an open problem. One potential way is to use parameterized two qubit rotation gates \cite{you2021exploring_VQE}.

Other possible future direction is to use efficient data preprocessing, for example reverse instance normalization \cite{kim2021reversible_instance_normalization} to mitigate the distribution shift between training and testing data. This is already being used in the state-of-the-art transformer based LTSF models like PatchTST \cite{nie2022_PatchTST} and MTST \cite{zhang2024multi_MTST}. These models also show better performance than the linear models in \cite{zeng2023_are_transformers_effective_for_time_series_forecasting}. Interestingly, our simple QuLTSF model outperforms or comparable to these models in limited settings. For instance, for the setting $(L=336, T=720)$ MSE of PatchTST, MTST and QuLTSF are $0.320$, $\underline{0.319}$ and $\mathbf{0.315}$ respectively. For the setting $(L=336, T=336)$ MSE of PatchTST, MTST and QuLTSF are $0.249$, $\mathbf{0.246}$ and $\underline{0.248}$ respectively (see Table $2$ in \cite{zhang2024multi_MTST} for PatchTST and MTST; and Table 1 for QuLTSF). QML based LTSF models with efficient data preprocessing may lead to improved results.

Implementation of QML based LTSF models on the quantum hardware poses significant challenges. As quantum systems grow, maintaining qubit state coherence and minimizing noise become increasingly difficult, leading to errors that degrade computational accuracy and performance \cite{book_lidar2013quantum_error_correction}. Additionally, the barren plateau phenomenon, where the gradient of the cost function vanishes as system size or circuit depth increases, further complicates optimization of VQC's \cite{mcclean2018barren_plateaus}. Addressing these challenges requires innovations in error correction, problem-specific circuit designs, and alternative optimization strategies, all of which are critical for enabling scalable, noise-resilient, and effective QML based LTSF models.

\section{\uppercase{Conclusions}}

We proposed \textit{QuLTSF}, a simple hybrid QML model for LTSF problems. QuLTSF combines the power of VQC's with classical linear neural networks to form an efficient LTSF model. Although simple linear models outperform more complex transformer-based LTSF approaches, incorporating a hidden quantum layer yielded additional improvements. This is demonstrated by extensive experiments on a widely used weather dataset showing QuLTSF’s superiority over the state-of-the-art classical linear models. This opens up a new direction of applying hybrid QML models for future LTSF research.

\section*{\uppercase{Acknowledgments}}

This research is sponsored by the Singapore Quantum Engineering Programme (QEP), project ID : NRF2021-QEP2-02-P06.


\bibliographystyle{apalike}
{\small
\bibliography{main}}


\end{document}